\documentclass[twocolumn,twoside,slac_two]{revtex4}
\usepackage{graphicx}
\usepackage{fancyhdr}
\newcommand{\ifb} {\ensuremath{\mathrm{fb^{-1}}}}

\newcommand{\ips} {\ensuremath{\mathrm{ps^{-1}}}}

\newcommand{\TeV}   {\ensuremath{\,\mathrm{Te\kern -0.1em V}}}
\newcommand{\GeVcc} {\ensuremath{\,\mathrm{Ge\kern -0.1em V\!/c}^2}}
\newcommand{\ds}    {D_s^-}
\newcommand{\phipi} {\phi\pi^-}
\newcommand{\ksk}   {K^{*0}K^-}
\newcommand{\ppp}   {\pi^+\pi^-\pi^-}
\newcommand{\Bd}    {\ifmmode B_d^0 \else $B_d^0$\fi}
\newcommand{\Bu}    {\ifmmode B^+ \else $B^+$\fi}
\newcommand{\Bs}    {\ifmmode B_s^0 \else $B_s^0$\fi}
\newcommand{\Bsb}   {\ifmmode \bar{B}_s^0 \else $\bar{B}_s^0$\fi}
\newcommand{\Bsh}   {\ifmmode B_s^H \else $B_s^H$\fi}
\newcommand{\Bsl}   {\ifmmode B_s^L \else $B_s^L$\fi}
\newcommand{\dgam}  {\ifmmode \Delta\Gamma \else $\Delta\Gamma$\fi}
\newcommand{\dms}   {\Delta m_s}
\newcommand{\dmd}   {\Delta m_d}
\newcommand{\lxy}   {\ifmmode L_{xy} \else $L_{xy}$\fi}
\newcommand{\pt}    {\ifmmode p_T \else $p_T$\fi}

\newcommand{\ld}    {\ifmmode \ell D \else $\ell D$\fi}
\newcommand{\Lb}    {\Lambda_b^0}
\newcommand{\Lc}    {\Lambda_c^+}
\newcommand{\Lzero} {\Lambda^0}

\begin{document}

\title{{\emph B} Mixing and Lifetimes at the Tevatron}


\author{G.~G\'omez-Ceballos}
\affiliation{Instituto de F\'{\i}sica de Cantabria (CSIC-UC),
  Avda. de los Castros s/n, 39005 Santander, Spain}

\author{J.~Piedra}
\affiliation{LPNHE-IN2P3/CNRS, Universites Paris VI et Paris VII,
%\affiliation{LPNHE-University Pierre et Marie Curie/CNRS-IN2P3,
  4 Place Jussieu Tour 33, 75252 Paris, France}

\begin{abstract}

The Tevatron collider at Fermilab provides a very rich environment for the study
of $b$-hadrons. Both the D\O~and CDF experiments have collected a sample of
about $1~\ifb$. We report results on three topics: $b$-hadron lifetimes,
polarization amplitudes and the decay width difference in $\Bs\to J/\psi\phi$,
and $\Bs$ mixing.

\end{abstract}

\maketitle

\thispagestyle{fancy}


\section{Introduction}
The Tevatron collider at Fermilab, operating at $\sqrt{s} = 1.96~\TeV$, has a
huge $b\bar{b}$ production cross section which is several orders of magnitude
larger than the production rate at $e^+e^-$ colliders running on the
$\Upsilon(4S)$ resonance. In addition, on the $\Upsilon(4S)$ only $\Bu$ and
$\Bd$ are produced, while higher mass $b$-hadrons such as $\Bs$, $B_c$,
$b$-baryons, $B^*$, and $p$-wave $B$ mesons are currently produced only at the
Tevatron.

Both D\O~and CDF~II are multipurpose detectors featuring high resolution
tracking in a magnetic field and lepton identification. These detectors are
symmetrical in polar and azimuthal angles around the interaction point, with
approximate $4\pi$ coverage~\cite{D0,CDF}. The CDF~II and D\O~detectors are able
to trigger at the hardware level on large track impact parameters, enhancing
the potential of their $B$ physics programs.

%-------------------------------------------------------------------------------
\section{Precision Lifetimes}
%-------------------------------------------------------------------------------

The lifetime of $b$-hadrons is governed primarily by the decay of the $b$-quark,
however contributions from the spectator quarks introduce small differences
between the lifetimes of different species. Presently these spectator effects
are mostly calculated in the framework of the Heavy Quark Expansion~\cite{tarantino}.

Several new results have been obtained by D\O~and CDF, most
of them already included in the world averages, and summarized in
Table~\ref{tab:lifetimes}~\cite{hfag}. The lifetime of the $\Lb$
baryon has been determined by both experiments through the
$\Lb \to J/\psi\Lzero$ decay,
\begin{eqnarray*}
\tau_{\Lb}^{D\O} =
1.22_{-0.18}^{+0.22}~(\mathrm{stat.}) \pm 0.04~(\mathrm{syst.})~\mathrm{ps}~[250~\mathrm{pb}^{-1}]\,,\\
\\
\tau_{\Lb}^{CDF} =
1.593_{-0.078}^{+0.083}~(\mathrm{stat.}) \pm 0.033~(\mathrm{syst.})~\mathrm{ps}~[1~\mathrm{fb}^{-1}]\,.
\end{eqnarray*}

\begin{figure}[hbt]
\vspace{9pt}
\includegraphics[width=15pc]{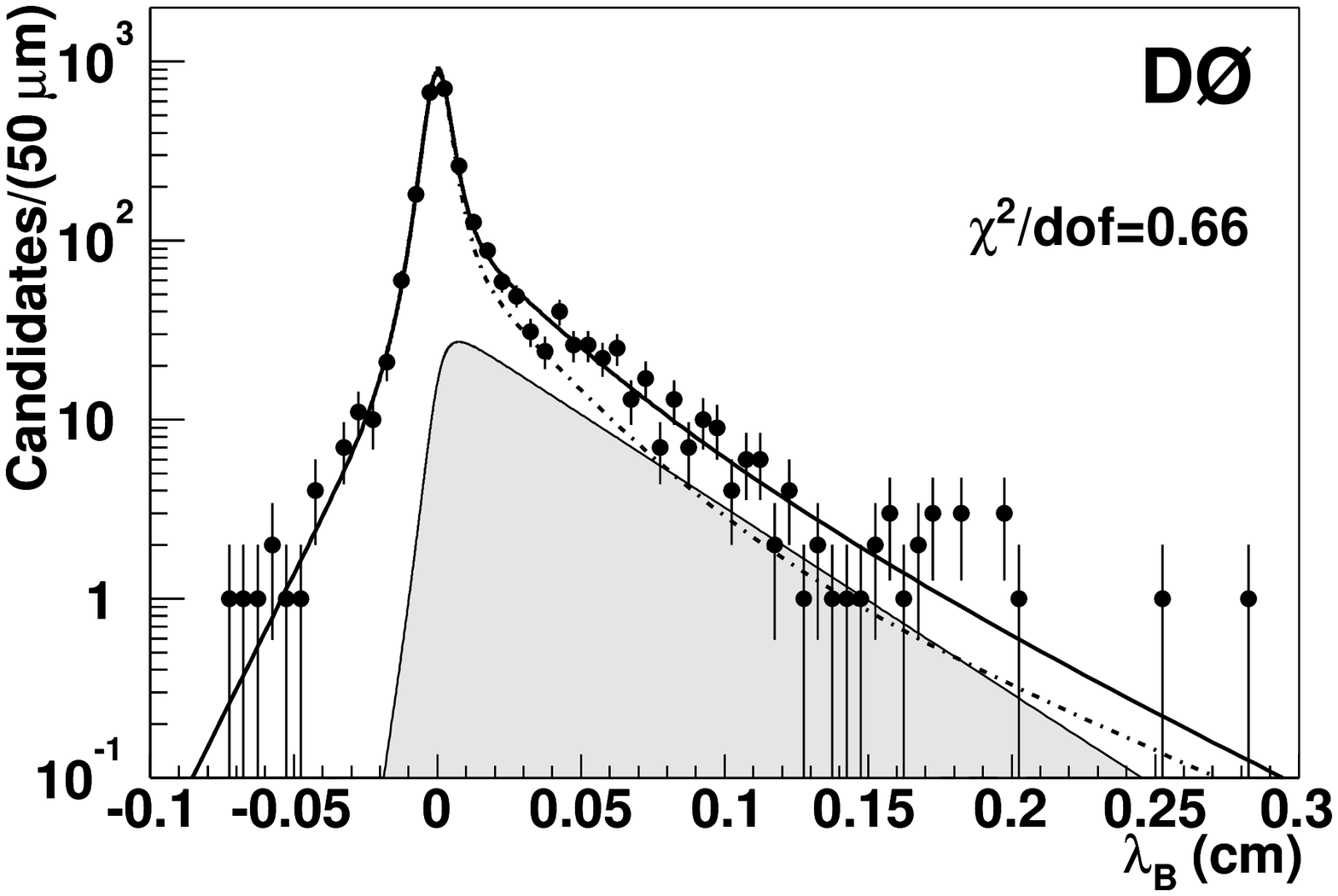}
\includegraphics[width=15pc]{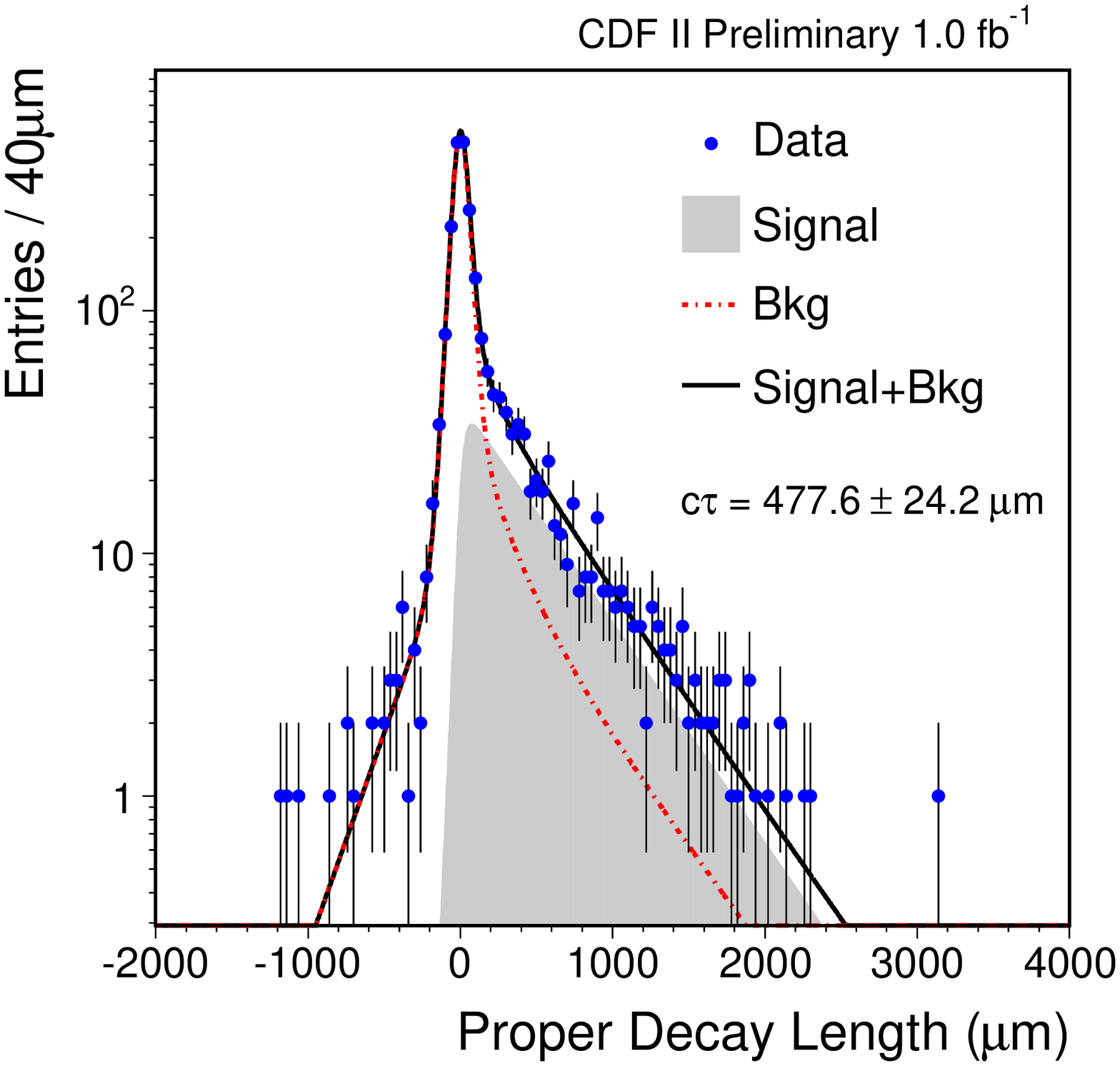}
\caption{Lifetime results from the $\Lb\to J/\psi\Lzero$ decay. The
points are the data, and the solid curve is the sum of fitted contributions
from signal (shaded area) and background (dot-dashed line).}
\label{fig:lb}
\end{figure}

Previous results used semileptonic $\Lb\to\Lc\ell^-\bar{\nu}_{\ell}$ decays.
Fully reconstructed decays are preferable because they do not suffer from
possible unexpected contributions from other $b$-baryon decays, and do not
require a correction for the missing momentum from unreconstructed decay
products ({\it{e.g.}}, the neutrino). The CDF measured value of
$\tau_{\Lb}$ is $3.1\sigma$ higher than the current world average
\cite{pdg}. This is the most precise single measurement of $\tau_{\Lb}$.
The fit projections of both lifetime measurements can be seen in
Fig.~\ref{fig:lb}.

In the case of the $\Bs$ meson, the lifetime is extracted from the semileptonic
decay $\Bs\to D_s^-\ell^+\nu$, providing the world's best measurement from D\O~\cite{d0bslife},
\begin{eqnarray*}
\tau_{\Bs}^{D\O} =
1.398 \pm 0.044~(\mathrm{stat.})_{-0.025}^{+0.028}~(\mathrm{syst.})~\mathrm{ps}~[400~\mathrm{pb}^{-1}]\,,
\\
\\
\tau_{\Bs}^{CDF} =
1.381 \pm 0.055~(\mathrm{stat.})_{-0.046}^{+0.052}~(\mathrm{syst.})~\mathrm{ps}~[370~\mathrm{pb}^{-1}]\,.
\end{eqnarray*}

\begin{table}[h]
\begin{center}
\caption{HFAG March 2006 averages compared to theory calculations.}
\begin{tabular}{|l|c|c|c|}
\hline
\textbf{$b$-hadron} & \textbf{lifetime} & \multicolumn{2}{c|}{\textbf{$\tau/\tau(\Bd)$}}\\
\textbf{species}    & \textbf{[ps]}     & \textbf{average} & \textbf{predicted range}\\
\hline
$\Bu$ & $1.643 \pm 0.010$ & $1.076 \pm 0.008$ & $1.04 - 1.08$\\
$\Bs$ & $1.454 \pm 0.040$ & $0.914 \pm 0.030$ & $0.99 - 1.01$\\
$\Lb$ & $1.288 \pm 0.065$ & $0.844 \pm 0.043$ & $0.86 - 0.95$\\
\hline
$\Bd$ & $1.527 \pm 0.008$ & --- & --- \\
$B_c$ & $0.469 \pm 0.065$ & --- & --- \\
\hline
\end{tabular}
\label{tab:lifetimes}
\end{center}
\end{table}

%-------------------------------------------------------------------------------
\section{Polarization Amplitudes}
%-------------------------------------------------------------------------------

The heavy $\Bsh$ and light $\Bsl$ mass eigenstates of the $\Bs$ meson are
mixtures of the two $CP$-conjugate states. Due to this mixture, the masses and
lifetimes of the mass eigenstates differ:
\begin{equation}
\Delta m\equiv m_H - m_L, \ \ \dgam\equiv\Gamma_L - \Gamma_H,
\end{equation}
where $m_{H,L}$ and $\Gamma_{H,L}$ denote the $B_s^{H,L}$ mass and decay width.
One of the important goals in Run~II is to measure the lifetime difference
$\Delta\Gamma$ between these two mass eigenstates. They are expected to be $CP$
eigenstates if the mixing phase is small. As it has been shown
theoretically~\cite{amplitudes_theory}, an angular analysis based on
transversity variables, combined with a lifetime measurement, permits one to
separate the $CP$-even and $CP$-odd final states of $\Bs\to J/\psi\phi$ (with
$J/\psi\to\mu^+\mu^-$ and $\phi\to K^+ K^-$), and hence determine the lifetime
difference.

Based on the single muon trigger, the D\O~experiment analyzes $0.8~\ifb$,
with $978 \pm 45$ $\Bs$ candidates passing the
selection cuts. A simultaneous unbinned likelihood fit is performed in terms of
invariant mass, proper decay length and transversity angular variables,
described in~\cite{pappas}. Due to limited detector coverage and kinematic
thresholds, the detector response to the transversity angles is non-uniform. The
acceptance is modelled with Monte Carlo simulation, reweighing the simulated
events to match the kinematic distributions observed in data.

In Fig.~\ref{fig:D0_Transversity} we show the projection of the fit result
onto the $\cos\theta$ transversity variable. Similar agreement is observed in
the projections onto the invariant mass, proper decay length and remaining
transversity angles. The results are presented in Table~\ref{tab:amplitudes}. The
$1\sigma$ contour for $\Delta\Gamma$ versus $c\bar{\tau}$, with
$\bar{\tau} = 1/\Gamma = 2/(\Gamma_H + \Gamma_L)$, is shown in Fig.~\ref{fig:D0_DGammaSummary}.
\begin{table}[h]
\begin{center}
\caption{Tevatron direct measurements of the decay rate difference between the
$\Bs$ mass eigenstates $\Delta\Gamma$, the average lifetime $\bar{\tau}$, the
fraction of the $CP$-odd component at $t=0$, $R_\perp = |A_\perp(0)|^2$, the
difference in quadrature of the $CP$-even linear polarization amplitude at $t=0$,
$|A_{0}(0)|^2 - |A_{\parallel}(0)|^2$, and the difference of the two
$CP$-conserving strong phases $\delta_1 - \delta_2$. The $CP$-violating weak
phase is assumed to be zero.}
\begin{tabular}{|l|c|c|}
\hline \textbf{observable} & \textbf{CDF~'04~\cite{dg_cdf}} & \textbf{D\O~'06~\cite{dg_d0}}\\
\hline
$\Delta\Gamma~[\ips]$                 & $0.47_{-0.24}^{+0.19}\pm0.01$ & $0.15_{-0.10-0.04}^{+0.10+0.03}$\\
$\bar{\tau}~[\mathrm{ps}]$            & $1.40_{-0.13}^{+0.15}$        & $1.53_{-0.08-0.04}^{+0.08+0.01}$\\
$R_\perp$                             & $0.13 \pm 0.08$               & $0.19 \pm 0.05 \pm 0.01$\\
$|A_{0}(0)|^2 - |A_{\parallel}(0)|^2$ & $0.355 \pm 0.067$             & $0.35 \pm 0.07 \pm 0.01$\\
$\delta_1 - \delta_2$                 & $1.94 \pm 0.36$               & $2.5 \pm 0.4$\\
\hline
\end{tabular}
\label{tab:amplitudes}
\end{center}
\end{table}

\begin{figure}[hbt]
\vspace{9pt}
\includegraphics[width=15pc]{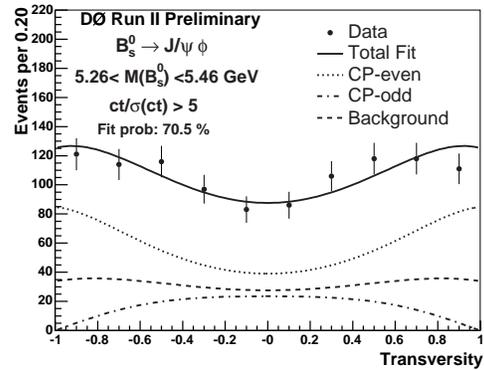}
\caption{$\cos\theta$ transversity distribution (D\O). The curves show: total
  fit (solid line), CP-even (dotted line), CP-odd (dot-dashed line) and
  background (dashed line).}
\label{fig:D0_Transversity}
\end{figure}

\begin{figure}[hbt]
\vspace{9pt}
\includegraphics[width=15pc]{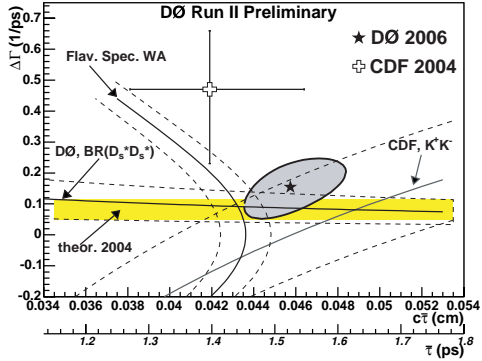}
\caption{The D\O~$1\sigma$ (stat.) contour in $\Delta\Gamma$ versus
  $c\bar{\tau}$ plane~\cite{dg_d0}, compared to a $1\sigma$ band for the world
  average based on flavor-specific decays~\cite{pdg}. The SM
  prediction~\cite{dg_sm} is shown as the horizontal band. Also shown are the
  CDF 2004 result~\cite{dg_cdf}, the recent CDF measurement of the $\Bs$
  lifetime from the $\Bs\to K^+ K^-$ decay~\cite{dg_kk}, and the implication of
  the D\O~result~\cite{D0_DsDs} of the branching fraction for the
  $\Bs\to D_s^{(*)+} D_s^{(*)-}$ decay.}
\label{fig:D0_DGammaSummary}
\end{figure}

%-------------------------------------------------------------------------------
\section{{\emph B} Mixing}
%-------------------------------------------------------------------------------

The mixing and $CP$ violation parameters of $B$ mesons are currently the focus
of much attention for pinning down the CKM matrix, and perhaps exposing new
physics beyond the Standard Model. The probability $\cal P$ for a $\Bs$ meson
produced at time $t=0$ to decay as $\Bs$ ($\Bsb$) at proper time $t>0$ is,
neglecting effects from $CP$ violation as well as a possible lifetime difference
between the heavy and light $\Bs$ mass eigenstates,
\begin{equation}
\label{Eq:prob_mix}
{\cal P}_{\pm} (t) =
\frac{\Gamma}{2} \, e^{- \Gamma t}\,
\left[ 1 \pm \cos ({\dms t} )\right],
\end{equation}
where the subscript ``$+$'' (``$-$'')
indicates that the meson decays as $\Bs$ ($\Bsb$). Oscillations have been
observed and well established in the $\Bd$ system. The mass difference~$\dmd$ is
measured to be~\cite{pdg}
\begin{eqnarray*}
  \dmd = 0.505 \pm 0.005~\ips\,.
\end{eqnarray*}

In the $\Bs$ system oscillations have also been well
established. Time-integrated measurements indicate that $\Bs$ mixing is large,
with a value of $\chi_s$~\cite{hfag} close to its maximal possible value of $1/2$,
\begin{eqnarray}
\chi_s &=& \frac{x_s^2 + y_s^2}{2\left(x_s^2 + 1\right)} >
0.49904~\mathrm{at~95\%~C.L.}\,,
\nonumber\\
2y_s &=& \frac{\Delta\Gamma}{\Gamma} = 0.31_{-0.11}^{+0.10}\,,
\\
x_s &=& \frac{\dms}{\Gamma} > 22.4~\mathrm{at~95\%~C.L.}\,\nonumber
\end{eqnarray}
However, the time dependence of this mixing has not been observed yet, and
previous attempts to measure $\dms$ have yielded a lower limit:
$\dms > 14.5~\ips$~\cite{pdg} at the 95\% confidence level (C.L.).

The canonical $B$ mixing analysis proceeds as follows. The $b$-flavor ($b$ or
$\bar{b}$) of the $B$ meson at the time of decay is determined from the charges
of the reconstructed decay products in the final state. The proper time at which
the decay ocurred is determined from the displacement of the $\Bs$ decay vertex
with respect to the primary vertex, and the $\Bs$ tranverse momentum with
respect to the proton beam. Finally, the production $b$-flavor must be known in
order to classify the $B$ meson as being mixed (production and decay $b$-flavor
are different) or unmixed (production and decay $b$-flavor are equal) at the
time of its decay.

%-------------------------------------------------------------------------------
\subsection{Signal Yields}
%-------------------------------------------------------------------------------

Both D\O~and CDF have performed mixing analyses using $1~\ifb$ of data
\cite{D0ANA,CDFANA}. The D\O~experiment exploits the high statistics single
muon trigger to study $\Bs\to\mu^+\ds X$, $\ds\to\phipi$ decays, reconstructing
26,700 signal candidates. On the CDF side, the analysis is performed
using both fully reconstructed $\Bs\to\ds (\pi^+\pi^-)\pi^+$ and 
semileptonic $\Bs\to\ell^+ \ds X$ ($\ell = e,\mu$) decays. In both cases 
the $\ds$ is reconstructed in the $\ds \rightarrow \phipi$, 
$\ds \rightarrow \ksk$ and $\ds \rightarrow \ppp$ modes. The signal yields
are 3,600 (fully reconstructed) and 37,000 (semileptonic).
Fig.~\ref{fig:D0_mass} (Fig.~\ref{fig:CDF_mass}) shows the reconstructed
candidates from D\O~(CDF).
\begin{figure}[htbp!]
\centering
\includegraphics[width=0.45\textwidth]{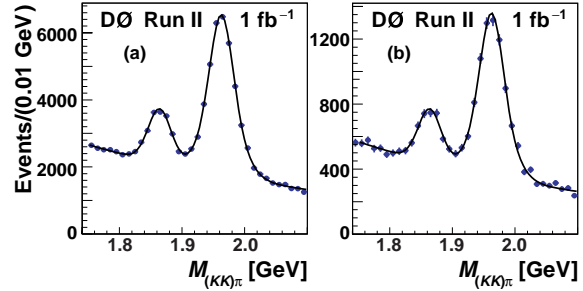}
\caption{$(K^+K^-)\pi^-$ invariant mass distribution for (a) untagged $\Bs$
sample, and (b) flavor-tagged $\Bs$ candidates. The left and right
peaks correspond to $\mu^+ D^-$ and $\mu^+\ds$ candidates, respectively.}
\label{fig:D0_mass}
\end{figure}

\begin{figure}[htbp!]
\centering
\includegraphics[width=0.45\textwidth]{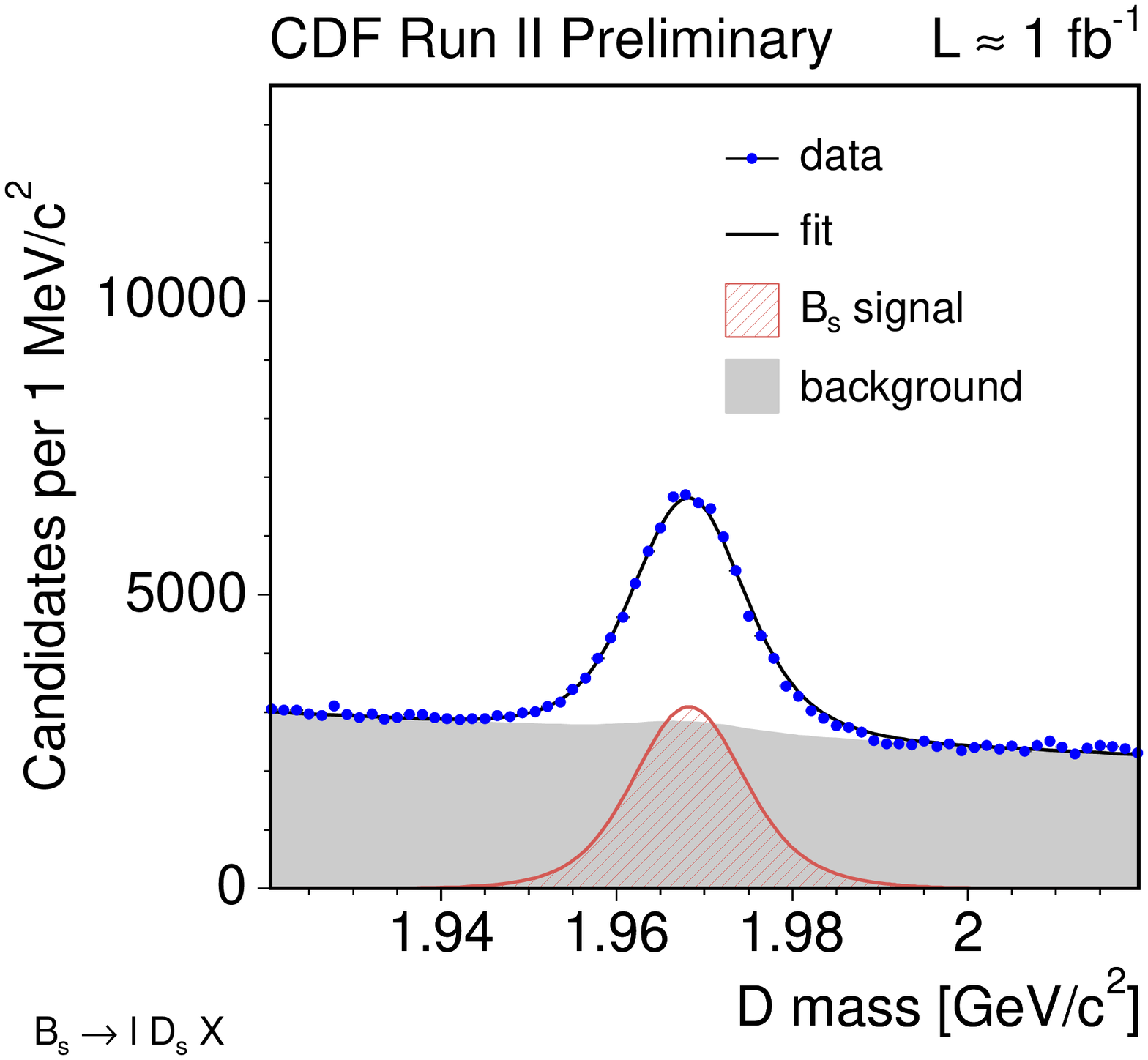}
\includegraphics[width=0.45\textwidth]{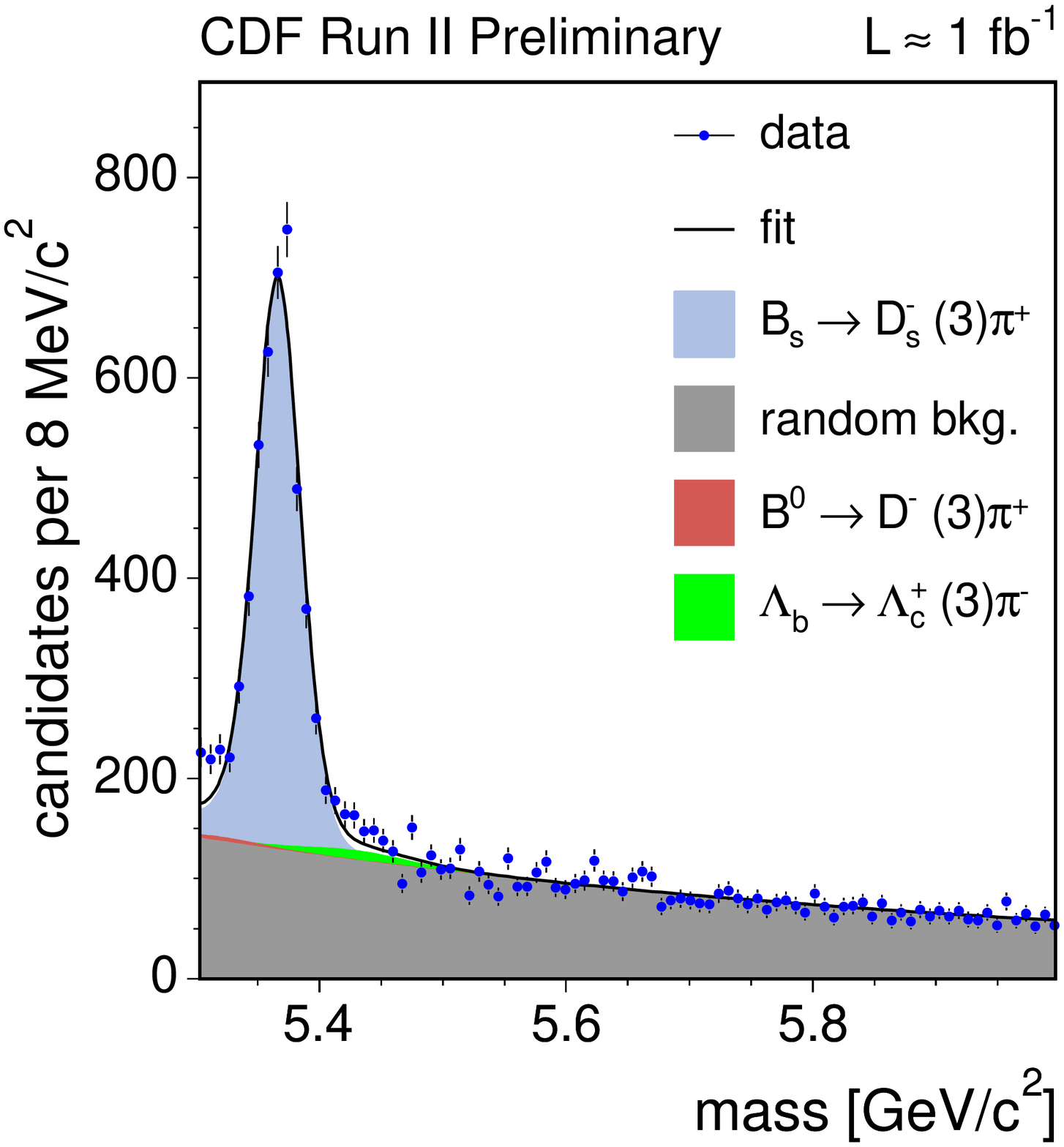}
\caption{$\ds$ invariant mass distribution for semileptonic decays (top)
and  $\Bs$ invariant mass distribution for fully reconstructed decays (bottom).}
\label{fig:CDF_mass}
\end{figure}

\subsection{Decay Length Reconstruction}
The transverse decay length $\lxy(B)$ is defined as the displacement from the
interaction point to the reconstructed $B$ vertex, in the plane transverse to
the proton beam. The $B$ meson decay time is then given by
%.......................................................................
\begin {equation}
t = \lxy(B) \, \frac{M_B}{\pt(B)}\,, 
\end{equation}
%.......................................................................
where $M_B$ is the nominal $B$ mass~\cite{pdg}. In semileptonic decays, the
$B$ meson is not fully reconstructed, and a correction factor has to be
included to account for the missing momentum,
%.......................................................................
\begin{equation}
t = \lxy(\ld) \, \frac{M_B}{\pt(\ld)} \,
k, \ \ k \equiv \frac{\lxy(B)}{\lxy(\ld)}\,\frac{\pt(\ld)}{\pt(B)}\,.
\end{equation}
%.......................................................................

The $k$-factor distribution is obtained from MC simulation. Both
D\O~and CDF use different $k$-factor distributions as a function of the $\ld$ mass,
as illustrated in Fig.~\ref{fig:kfactor_vs_mld}.

\begin{figure}[htb!]
\begin{center}
\includegraphics[width=0.45\textwidth]{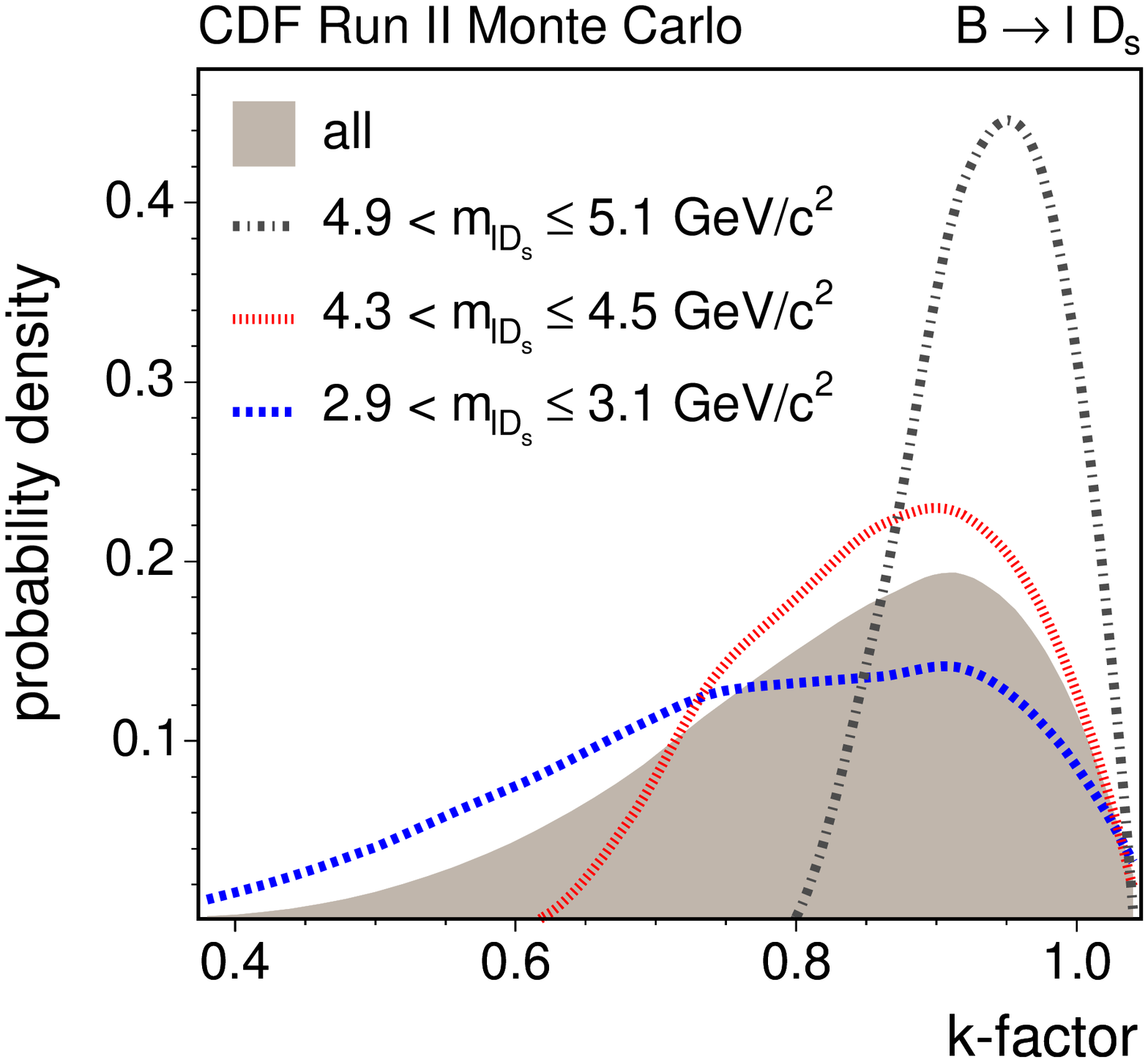}
\caption{$k$-factor distribution for several $\ld$ mass regions, for 
$\Bs \to \ell^+ \ds X, \ds \to \phipi$ decays.}
\label{fig:kfactor_vs_mld}
\end{center}
\end{figure}

%% EFFICIENCY CURVE
%% -- %% Due to some trigger and selection requirements related to the reconstructed
%% -- %% proper decay time distribution, the proper decay time does not follow a pure
%% -- %% exponential (modulo resolution and $k$-factor effects), but it is biased. This
%% -- %% bias, expressed as an efficiency curve, is obtained using Monte Carlo simulation.

%% CT RESOLUTION
%% -- %% The determination of the $ct$ resolution is crucial for the derivation 
%% -- %% of a proper result on $\dms$.
%% -- %% CDF uses unbiased prompt events to measure $ct/\sigma_{ct}$, and
%% -- %% then applies an event-by-event correction, which depends on the topology and on 
%% -- %% several kinematical quantities. D\O~uses a sample of $J/\psi \to \mu^+ \mu^-$ decays to
%% -- %% examine the $ct/\sigma_{ct}$ distribution, and then applies an overall
%% -- %% correction assuming a double Gaussian distribution.

The determination of the time resolution is a necessary piece of input for derivation 
of a proper result on $\dms$. The most precise determinations come from lifetime 
measurements in the exclusively reconstructed modes involving $J/\psi$ decays. 
Usually, the time resolution is part of the fitting procedure and is determined in 
the $B$ meson sidebands, which contain a large fraction of prompt events. In those prompt 
$J/\psi$ candidates, another track from the primary vertex is combined with the 
$J/\psi$, and is ideally suited for measuring the resolution on $t$ with respect to 
the primary vertex position. For analyses which rely on samples 
obtained with a selection that biases the time distribution, these prompt components in the
sidebands are greatly reduced, which makes the determination of the time resolution 
problematic. CDF uses unbiased prompt events to measure $t/\sigma_{t}$, and
then applies an event-by-event correction, which depends on the decay topology and on 
several kinematical quantities. D\O~uses a sample of $J/\psi \to \mu^+ \mu^-$ decays to
examine the $t/\sigma_{t}$ distribution, and then applies an overall
correction on the time resolution with a double Gaussian distribution: the narrow Gaussian has a width 
of $0.998\sigma$ and comprises 72\% of the total, and the second Gaussian has a 
width of $2.777\sigma$, where $\sigma$ is the estimated error on the 
proper decay time of the candidate.

\subsection{Flavor Tagging}
The methods of $b$-flavor tagging may be classified into two categories:
opposite-side (soft lepton and jet charge) and same-side production $b$-flavor identification.
Opposite-side taggers exploit the fact that $b$ quarks in hadron colliders are
mostly produced in $b\bar{b}$ pairs. Same-side flavor tags are based on the
charge of particles produced in association with the production of the
$b$-hadron.

The performance of the $b$-flavor taggers is quantified by their
efficiency $\epsilon$ and dilution ${\cal D} = 2P_{tag} - 1$, where $P_{tag}$
is the probability for the production $b$-flavor to be correctly identified.
For setting a limit on $\dms$, knowledge of the flavor taggers performance is
crucial. 

\subsection{Opposite-Side Flavor Tagging}
\label{sec:ost}

The soft lepton tagger (SLT) is based on semileptonic $b$ decays into an
electron or a muon ($b\rightarrow \ell^-X$). The charge of the lepton is
correlated to the charge of the decaying $B$ meson. The jet charge tagger (JQT)
uses the fact that the charge of a $b$-jet is correlated to the charge of the
$b$ quark.

The performance of the opposite-side flavor taggers is measured in kinematically
similar $\Bd$ and $\Bu$ samples, and we summarize it in Table~\ref{tab:tagger}.
The analysis involves complex fits combining several $\Bd$ and $\Bu$ decay
modes. Both D\O~and CDF have measured
$\dmd$ and find
\begin{eqnarray*}
\dmd^{D\O} =
0.506 \pm 0.020~(\mathrm{stat.}) \pm 0.016~(\mathrm{syst.})~\ips\,,\\
\\
\dmd^{CDF} =
0.509 \pm 0.010~(\mathrm{stat.}) \pm 0.016~(\mathrm{syst.})~\ips\,.\\
\end{eqnarray*}

\begin{table}[h]
\begin{center}
\caption{Opposite-side flavor taggers performance.}
\label{tab:tagger}
\begin{tabular}{|l|c|c|}
\hline
& \multicolumn{2}{|c|}{$\epsilon {\cal D}^2$ [\%]} \\
\hline \textbf{tagger} & \textbf{D\O} & \textbf{CDF}\\
\hline
 muon                & $1.48 \pm 0.17$ & $0.55 \pm 0.05$ \\
 electron            & $0.21 \pm 0.07$ & $0.30 \pm 0.03$ \\
 JQT                 & $0.50 \pm 0.11$ & $0.70 \pm 0.06$ \\
\hline
 combined            & $2.48 \pm 0.22$ & $1.55 \pm 0.08$ \\
\hline
\end{tabular}
\end{center}
\end{table}

\subsection{Same-Side Flavor Tagging}
\label{sec:sst}
During the fragmentation and the formation of the $\Bs$ meson there is a 
left over $\bar{s}$ quark which may form a $K^+$. Hence,
if there is a nearby charged particle, which is additionally identified as a 
kaon, it is quite likely that it is the leading fragmentation track and its 
charge is then correlated to the flavor of the $\Bs$ meson. While the 
performance of an opposite-side tagger does not depend on the flavor of 
the $B$ on the signal side, the same-side tagger performance depends on the signal 
fragmentation process. Therefore the opposite-side performance can be 
measured in $\Bd$ and $\Bu$ samples, and can then be used for setting a limit on the 
$\Bs$ mixing frequency. But when using a same-side tagger for a limit on $\dms$, 
one must rely on Monte Carlo simulation.

CDF has performed extensive data and Monte Carlo comparisons on several quantities
related to the tagging, and determined the tagging candidate by selecting the most 
likely kaon track. To separate kaons from other particle species, a combined
particle identification likelihood based on information from $dE/dx$ and
from the Time-of-Flight system has been used. A comparison 
between data and {\tt{PYTHIA}} Monte Carlo~\cite{PYTHIA} for the dilution
obtained by using that likelihood is shown in Fig.~\ref{fig:cll_dil}. There we
can see an excellent agreement between the results
obtained from Monte Carlo, and those from data, thus providing confidence on
the performance description in Monte Carlo. The
effectiveness of this flavor tag increases with the $p_T$ of the $\Bs$. The
values found using {\tt{PYTHIA}} Monte Carlo are $\epsilon{\cal D}^2 =
3.5\%~(4.0\%)$ in the hadronic (semileptonic) decay sample.

\begin{figure}[htb!]
\begin{center}
\includegraphics[width=0.45\textwidth]{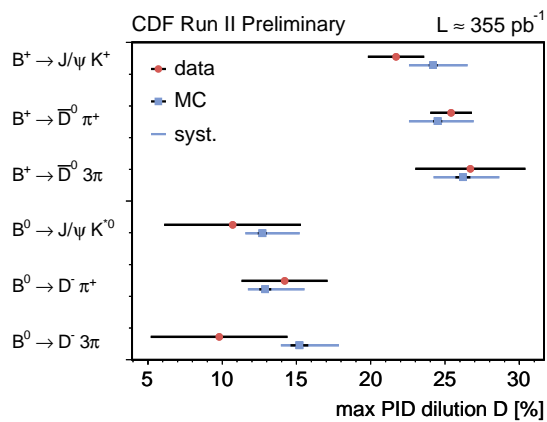}
\caption{Comparison between CDF data and {\tt{PYTHIA}} Monte Carlo for the dilution
obtained by selecting the most likely kaon track as the tag.}
\label{fig:cll_dil}
\end{center}
\end{figure}

\subsection{Amplitude Scan}
The likelihood term describing the proper decay time of flavor-tagged neutral
$B$ meson candidates is modified by including an additional
parameter multiplying the cosine, the so-called amplitude ${\cal A}$,
\begin{equation}
{\cal L}_{\mathrm{signal}} \propto 1\pm {\cal A} \, {\cal D} \cos(\Delta m t)\,.
\end{equation}

The parameter ${\cal A}$ is left free in the fit while ${\cal D}$ is known, as
explained in Sec.~\ref{sec:ost} and~\ref{sec:sst}, and
fixed in the scan. This method~\cite{MOSER} involves performing one
such ${\cal A}$-fit for each value of the parameter $\Delta m$, which is fixed
at each step; in the case of infinite statistics, optimal resolution and perfect
dilution calibration, one would expect ${\cal A}$ to be
unity for the true oscillation frequency and zero for the remainder of
the probed spectrum. In practice,
the output of the procedure is accordingly a list of fitted values
(${\cal A}$, $\sigma_{\cal A}$) for each $\Delta m$ hypothesis.
A particular $\Delta m$ hypothesis is excluded to a $95$\% confidence
level in case the following relation is observed:
${\cal A} + 1.645 \,\sigma_{\cal A} < 1$. The sensitivity of a mixing measurement
is defined as the $\Delta m$ value for which $1.645 \,\sigma_{\cal A} = 1$.

The scan shown in Fig.~\ref{fig:dmd_amplitude} is obtained when the method is
applied to $\Bd\to J/\psi K^{*0}$ and $\Bd\to D^-\pi^+$ samples from the CDF
experiment, using the combined opposite-side tagging algorithms. The expected
compatibility of the measured amplitude with unity in the vicinity of the true
frequency, $\dmd = 0.5~\ips$, is verified.

\begin{figure}[htb!]
\begin{center}
\includegraphics[width=0.45\textwidth]{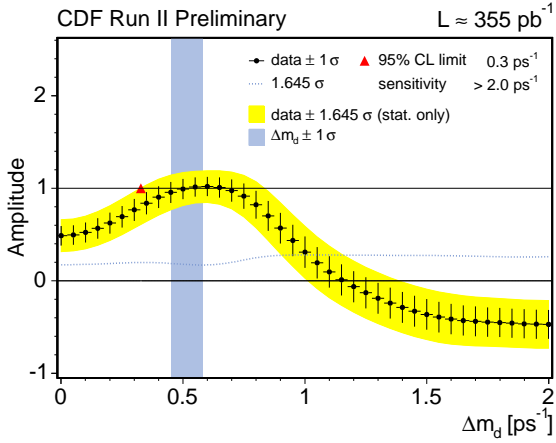}
\caption{$\Delta m_d$ ${\cal A}$-scan in hadronic decays (CDF).}
\label{fig:dmd_amplitude}
\end{center}
\end{figure}

\subsection{D\O~\boldmath$\dms$ Results}
The D\O~amplitude scan on $1~\ifb$ is shown in Fig.~\ref{fig:combinedscan_d0}.
The sensitivity is 14.1~$\ips$, and the 
95\% C.L. limit is $\dms>14.8$~$\ips$. Fig.~\ref{fig:likelihood_d0} shows
the ratio of the likelihood function at ${\cal A}=0$ and ${\cal A}=1$.
The preferred value assuming a signal is $\dms$ = 19 $\ips$,
with a 90\% C.L. interval of 17 $<~\dms~<~$ 21 $\ips$. The probability that the
random-tag background could fluctuate to mimic such a signature is about 5\%. 

\begin{figure}[htb!]
\begin{center}
\includegraphics[width=0.45\textwidth]{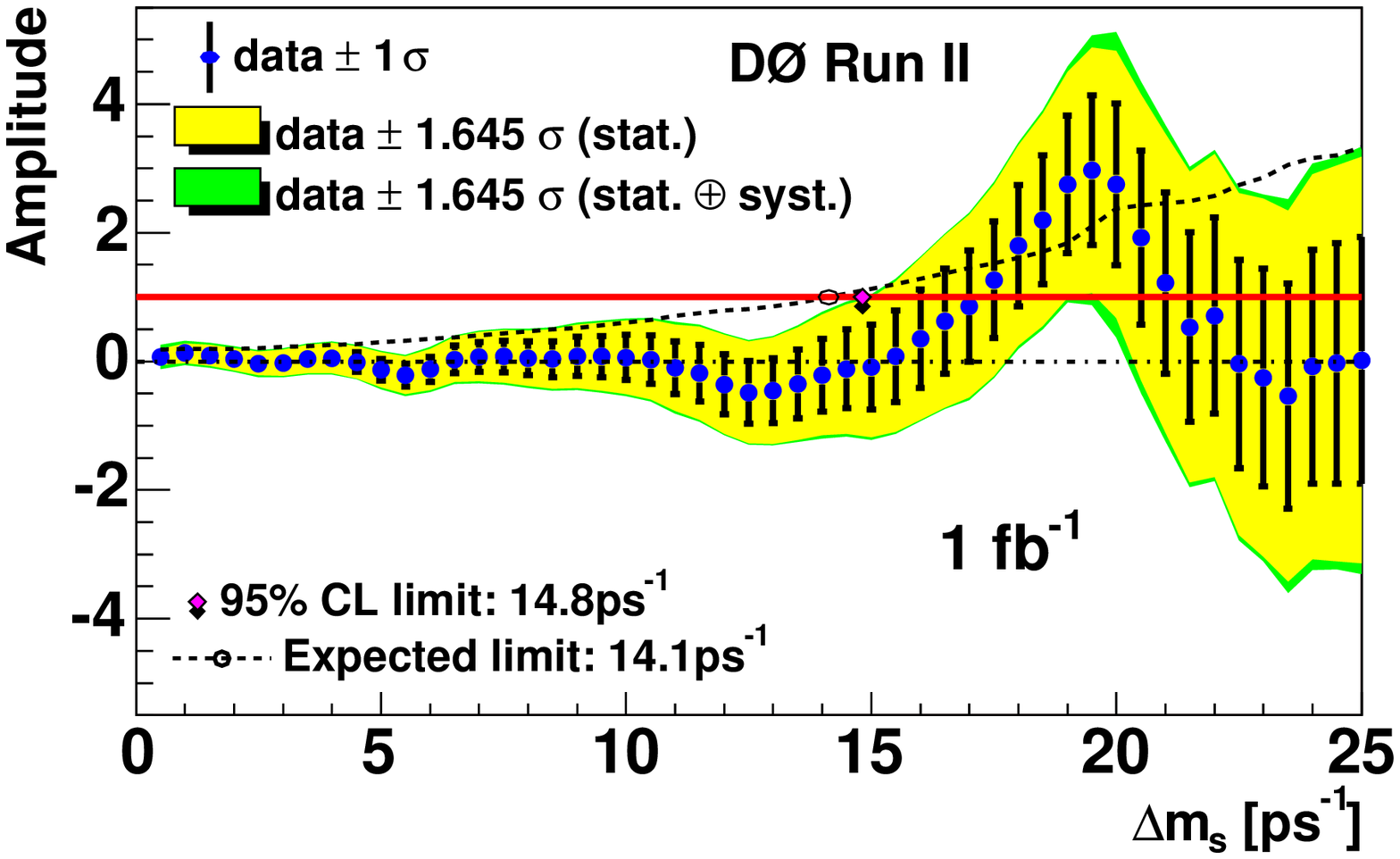}
\caption{$\dms$ ${\cal A}$-scan (D\O). The dotted line represents
$1.645\,\sigma_{\cal A}$, indicating a sensitivity of $\dms = 14.1~\ips$.}
\label{fig:combinedscan_d0}
\end{center}
\end{figure}

\begin{figure}[htb!]
\begin{center}
\includegraphics[width=0.45\textwidth]{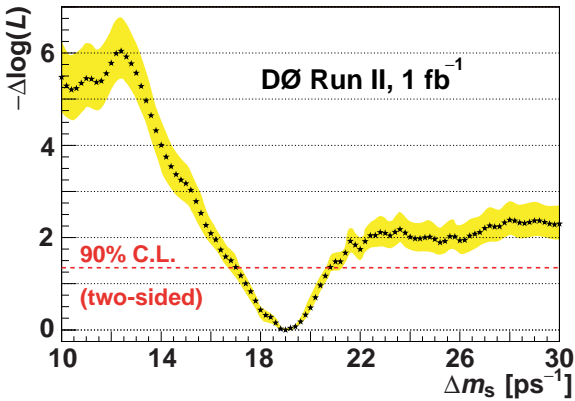}
\caption{Value of -$\Delta\cal{L}$ as a function of $\dms$ (D\O). The shaded
  band represents the envelope of all $\log{\cal L}$ scan curves due to
  different systematic uncertainties.}
\label{fig:likelihood_d0}
\end{center}
\end{figure}

\subsection{CDF~\boldmath$\dms$ Results}
The CDF combined amplitude scan on $1~\ifb$ is shown in 
Fig.~\ref{fig:combinedscan_cdf}. The sensitivity for the combination of all 
hadronic and semileptonic modes is $25.8~\ips$, and the 95\% C.L. limit is 
$\dms>16.7~\ips$.

\begin{figure}[htb!]
\begin{center}
\includegraphics[width=0.45\textwidth]{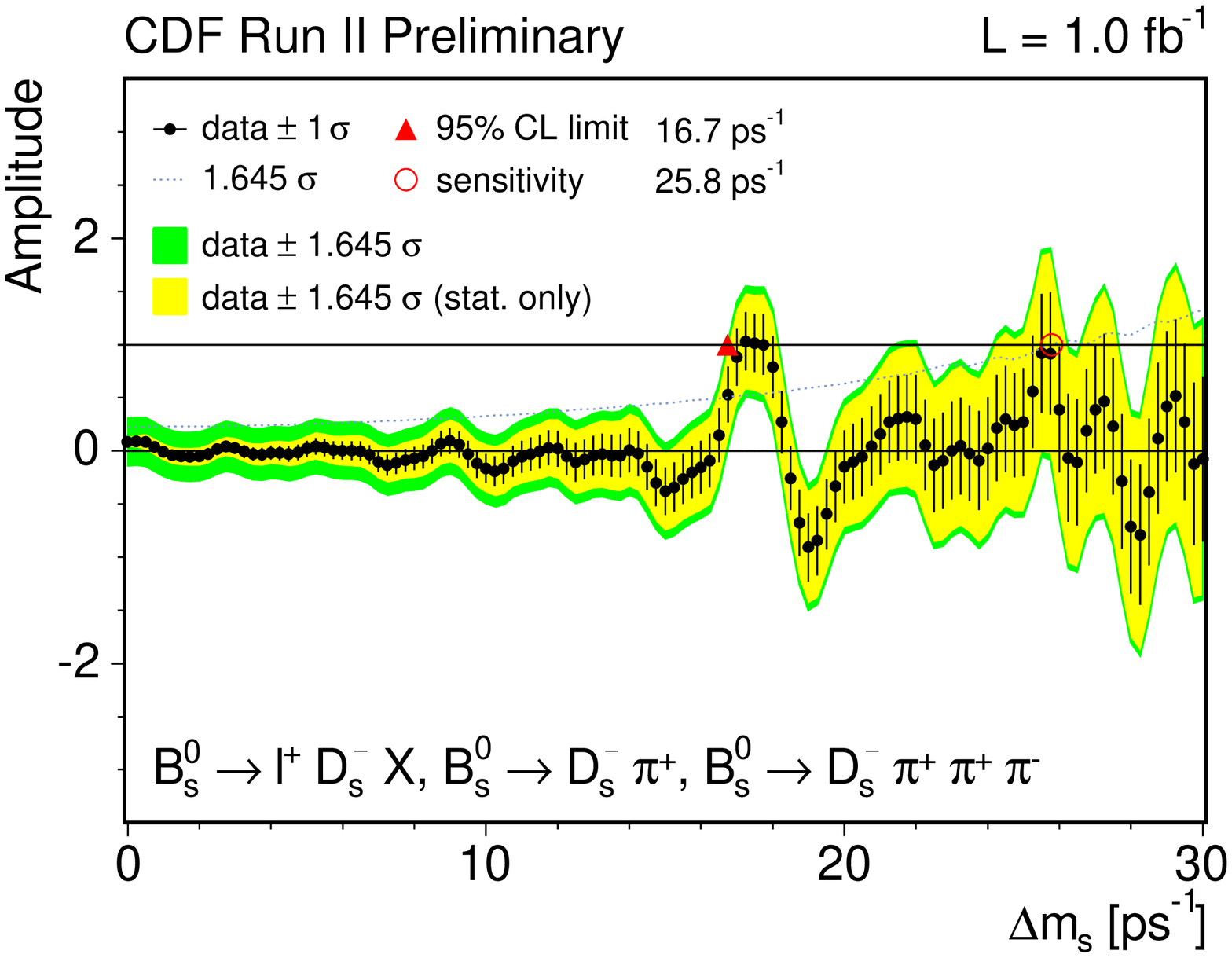}
\caption{$\dms$ ${\cal A}$-scan (CDF). The dotted line represents
  $1.645\,\sigma_{\cal A}$, indicating a sensitivity of $\dms = 25.8~\ips$.}
\label{fig:combinedscan_cdf}
\end{center}
\end{figure}

The amplitude shows a value consistent with unity near $\dms = 17.3~\ips$. 
To assess the significance of this deviation, CDF looked at the ratio of the 
likelihood function at ${\cal A}=0$ and ${\cal A}=1$, as shown in 
Fig.~\ref{fig:likelihoodratio}. The maximum likelihood ratio is at 
$\dms = 17.3~\ips$ and has an absolute value of 6.75. The probability that
the random-tag background could fluctuate to mimic such a signature is 0.2\%. Under the hypothesis that 
this is a signal for $\Bs-\bar{B}_s^0$ oscillations, CDF measures
\begin{eqnarray*}
\dms &=& 17.31_{-0.18}^{+0.33}~(\mathrm{stat.})\pm 0.07~(\mathrm{syst.})~\ips\,,
\end{eqnarray*}
with the systematic error completely dominated by the time scale uncertainty,
which is $0.4\%$.

\begin{figure}[htb!]
\begin{center}
\includegraphics[width=0.45\textwidth]{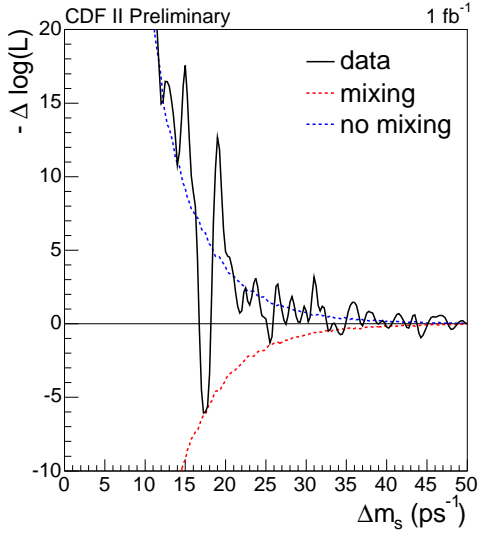}
\caption{Combined likelihood ratio as a function of $\dms$ (CDF).}
\label{fig:likelihoodratio}
\end{center}
\end{figure}

\section{Conclusions}
Lifetime measurements have been made in the clean $\Lb\to J/\psi\Lzero$
decay. The CDF result is the most precise measurement of $\tau(\Lb)$, and
the first using a fully reconstructed mode that reaches a precision comparable
with the previous best measurements based upon semileptonic decays of the
$\Lb$. Both D\O~and CDF have measured the width difference between the light
and heavy $\Bs$ mass eigenstates, which, in the limit of no $CP$ violation,
coincide with the $CP$-even and $CP$-odd eigenstates of the $\Bs$ system.

D\O~has performed a study of $\Bs-\Bsb$ oscillations using $\Bs \to \mu^+ \ds X$
decays and opposite-side flavor tagging algorithms. The expected limit at
95\% C.L. is 14.1 $\ips$. Assuming Gaussian uncertainties, a 90\% C.L. interval of
$17 < \dms < 21~\ips$ is set.

CDF has searched for $\Bs$ flavor oscillations using hadronic and semileptonic
decays. Opposite-side and, for the first time at a hadron collider, same-side
tags provide information about the \Bs~production flavor. A significant peak in
the amplitude scan consistent with unity is observed. Assuming this is a signal
for $\Bs-\Bsb$ oscillations, CDF measures
\begin{eqnarray*}
\dms &=& 17.31_{-0.18}^{+0.33}~(\mathrm{stat.})\pm 0.07~(\mathrm{syst.})~\ips\,.
\end{eqnarray*}

The $\Bs-\Bsb$ oscillation frequency measured at CDF is used to derive the ratio
$|V_{td}/V_{ts}|$,
\begin{eqnarray*}
\left|\frac{V_{td}}{V_{ts}}\right| &=&
\xi \sqrt{\frac{\dmd M_{\Bs}}{\dms M_{\Bd}}}\\
&=& 0.208_{-0.002}^{+0.001}~(\mathrm{exp.})_{-0.006}^{+0.008}~(\mathrm{theo.})\,,
\end{eqnarray*}
where the following values have been used as inputs: $M_{\Bd}/M_{\Bs} = 0.98390$
\cite{pdg} with negligible uncertainty, $\dmd = 0.505 \pm 0.005~\ips$~\cite{pdg}
and $\xi = 1.21_{-0.035}^{+0.047}$~\cite{lattice05}.

\bigskip % extra skip inserted
\begin{acknowledgments}

We would like to thank the Local Organizing Committee for a very enjoyable
conference, and for accomodating the request for an extra talk on $\Bs$
mixing at CDF.

J.~Piedra is supported by the EU funding under the RTN contract:
HPRN-CT-2002-00292, Probe for New Physics.

\end{acknowledgments}

\bigskip % extra skip inserted
%\begin{thebibliography}{9}   % Use for  1-9  references

\end{document}